\documentclass[12pt,epsf,amssymb]{article}
      \def\lsim{\raise0.3ex\hbox{$<$\kern-0.75em\raise-1.1ex\hbox{$\sim$}}}
\def\gsim{\raise0.3ex\hbox{$>$\kern-0.75em\raise-1.1ex\hbox{$\sim$}}}
\def\noi{\noindent}
\def\nn{\nonumber}
\def\bea{\begin{eqnarray}}  \def\eea{\end{eqnarray}}
\def\beq{\begin{equation}}   \def\eeq{\end{equation}}

\def\beeq{\begin{eqnarray}} \def\eeeq{\end{eqnarray}}
\def\R{ {\rm R \kern -.31cm I \kern .15cm}}
\def\C{ {\rm C \kern -.15cm \vrule width.5pt \kern .12cm}}
\def\Z{ {\rm Z \kern -.27cm \angle \kern .02cm}}
\def\N{ {\rm N \kern -.26cm \vrule width.4pt \kern .10cm}}
\def\1{{\rm 1\mskip-4.5mu l} }

\usepackage{graphics}
\usepackage{graphicx}
\usepackage{epsfig}

\begin{document}
\begin{center}
{\large \bf Bounds on the derivatives of the Isgur-Wise function with 
a non-relativistic light quark} \\

\vskip 1 truecm
{\bf F. Jugeau, A. Le Yaouanc, L. Oliver and J.-C. Raynal}\\
 
{\it Laboratoire de Physique Th\'eorique}\footnote{Unit\'e Mixte de Recherche
UMR 8627 - CNRS }\\    {\it Universit\'e de Paris XI, B\^atiment 210, 91405
Orsay Cedex, France}
\end{center}

\vskip 1 truecm

\begin{abstract} In a preceding study in the heavy quark limit of 
QCD, it has been
demonstrated that the best lower bound on the curvature of the
Isgur-Wise function $\xi (w)$ is $\xi '' (1) > {1 \over 5} [4\rho^2 +
3(\rho^2)^2] > {15 \over 16}$. The quadratic term $(\rho^2)^2$ is
dominant in a non-relativistic expansion in the light quark, both $\xi
'' (1)$ and $(\rho^2)^2$ scaling like $(R^2m_q^2)^2$, where $m_q$ is
the light quark mass and $R$ the bound state radius. The
non-relativistic limit is thus a good guide-line in the study of the
shape of $\xi (w)$. In the present paper we obtain similar bounds on all the
derivatives of $\xi_{NR}(w)$, the IW function with the light quark
non-relativistic, and we demonstrate that these bounds are
optimal. Our general method is based on the
positivity of matrices of moments of the ground state wave function,
that allows to bound the $n$-th derivative $\xi_{NR}^{(n)}(w)$ in terms of the
$m$-th ones ($m < n$). We show that the method can be generalized to the true
Isgur Wise function of QCD $\xi (w)$. \end{abstract}

\vskip 2 truecm

\noi LPT Orsay 04-14 \par
\noi February 2004\par \vskip 1 truecm

\noindent e-mails : frederic.jugeau@th.u-psud.fr, 
leyaouan@th.u-psud.fr, oliver@th.u-psud.fr
\newpage
\pagestyle{plain}

\section{Introduction.}
\hspace*{\parindent}
Using the OPE in the heavy quark limit of QCD, new Bjorken-like sum
rules (SR) have been obtained \cite{1r,2r,3r}. It has been shown that
the Isgur-Wise (IW) function $\xi (w)$ is an alternate series in powers
of $(w-1)$, and lower bounds have been found on the absolute magnitude
of its derivatives. Important ingredients in the derivation of the SR
are the consideration, following Uraltsev \cite{4r} of the non-forward
amplitude, plus the systematic use of boundary conditions that ensure
that only a finite number of $j^P$ intermediate states (with their
tower of radial excitations) contribute.\par

In particular, it has been found that the $n$-th derivative is bounded
by the $(n-1)$-th one \cite{2r}

\beq \label{1e} (-1)^n \xi^{(n)}(1) \geq {2 n + 1 \over 4} \ (-1)^{n-1}
\xi^{(n-1)} (1) \geq {(2n+1)!! \over 2^{2n}}\eeq

\noi where the second inequality follows from the recursive character
of the first one, and generalizes the inequality for the slope $\rho^2
= - \xi '(1)$~:

\beq \label{2e} \rho^2 \geq {3 \over 4} \eeq

\noi that follows from Bjorken \cite{5r} and Uraltsev \cite{4r} SR. The
first inequalities (\ref{1e}) read, for the curvature $\sigma^2 = \xi ''
(1)$

\beq \label{3e} \sigma^2 \geq {5 \over 4} \ \rho^2 \geq {15 \over 16} \ .\eeq

In \cite{3r} one has obtained, from a wider class of SR, the following
better bound on the curvature

\beq \label{4e} \sigma^2 \geq {1 \over 5} \left [ 4 \rho^2 + 3(\rho^2)^2
\right ] \geq {15 \over 16} \eeq

\noi where the absolute lower bound (independent of $\rho^2$) follows
from (\ref{2e}). Radiative corrections to the bounds (\ref{3e}) and 
(\ref{4e}) have been computed by M. Dorsten \cite{6new}.\par

It has been underlined in \cite{3r} that the quadratic term in
(\ref{4e}) has a clear physical interpretation, as it is leading in a
non-relativistic (NR) expansion in the mass of the light quark~:

\beq \label{5e} \xi ''_{NR}(1) > {3 \over 5} (\rho^2)^2 \eeq

\noi where $\xi_{NR}(w)$ denotes the Isgur-Wise function with the light
quark in the NR limit.\par

It is clear that it is very important to have rigorous bounds on the
derivatives of the IW function. The main general reason is that the
shape of the latter is linked to the determination of the CKM matrix
element $|V_{cb}|$ through the exclusive processes $B \to D^{(*)}\ell
\nu$. As pointed out in \cite{3r}, a more quantitative reason is that,
beyond the first derivative, higher derivatives will play a
non-negligible role at the edge of the phase space, at high $w$, in the
region where the data are presently rather precise, and will become
more and more precise in the near future.\par

Therefore, it is not only of an academic interest to find bounds on
higher derivatives. To this aim, one can begin by using systematically
for higher derivatives the method exposed in \cite{3r}. It is then
possible to find bounds of the form (\ref{4e}) for higher derivatives
\cite{6r}. \par

However, we have realized that, studying the NR situation, a more
powerful method can be developed that leads to better bounds,
exposed below. With this method, one finds for the second derivative,
in the NR limit, the bound (\ref{5e}), but for higher derivatives
$\xi_{NR}^{(n)}(1)$ one finds more complicated bounds involving the $m$-th
ones ($m < n$). It is important to notice that, unlike the method 
that we have used in
QCD \cite{1r}-\cite{3r}, where the IW functions to excited states played
a crucial role, in the present paper we note that a simple general
property (positivity) involving only the elastic IW function allows to
deduce all the bounds on its derivatives at zero recoil.\par

In QCD, in the heavy quark limit one can hope to obtain bounds such
that, in the NR limit for the light quark, reduce to these better
bounds. For the moment, the aim of the present paper is to obtain
bounds for the derivatives of $\xi_{NR}(w)$, opening the way to study
the more complex case of the actual IW function $\xi (w)$ in QCD. \par

In Section 2 we set the relation between the derivatives
$\xi_{NR}^{(m)}(1)$ and the moments $<0|z^m|0>$, where $|0>$ is the
ground state radial wave function. In Section 3 we study the general
constraints on the moments. In Section 4 we shift to the corresponding
constraints on the derivatives and the resulting bounds. In Section 5
we illustrate our bounds by particularizing to simple potentials and in
Section 6 we conclude. In Appendix A we demonstrate a mathematical
identity used in the text, in Appendix B we show the frame dependence
of subleading moments and in Appendix C we demonstrate the optimality
of the bounds. In Appendix D we deduce a simple formula giving a weaker
but completely explicit bound for all the even derivatives
$\xi^{(2n)}(1)$, that we compare with our optimal bounds.

\section{Isgur-Wise function in the NR limit : relation between its 
derivatives and moments.}
\subsection{Universal NR form factors in heavy-heavy transitions.}
\hspace*{\parindent}
As it is well known from nuclear physics, the first-quantized
non-relativistic operator corresponding to the electromagnetic current
$e_Q\overline{Q}(x)\gamma^0Q(x)$ of field theory is given by the
expression $e_Q \delta ({\bf x} - {\bf r}_Q)$, where ${\bf r}_Q$ is the
position of the active quark (see for example \cite{7newr}). Here we 
are interested in the form factor corresponding to a general
current $\overline{Q}\, '(x) \Gamma Q(x)$ for transitions between bound
states of the type $(Q,q) \to (Q',q)$ with unequal masses. For
simplicity, let us begin with the matrix element of the fourth component of the vector current
$J^0(0) = \overline{Q}\, '(0) \gamma^0 Q(0)$. \par

Let us write the corresponding non-relativistic (NR) form factor in a 
general
frame. The form factor for the transition $(Q,q) \to (Q',q)$
will be given simply by the matrix element of the operator 
$\delta({\bf r}_Q)$~:

\beq \label{35e} F({\bf P'} , {\bf P}) = < {\bf P'} |\delta({\bf
r}_Q)|{\bf P}> = \int \Psi_{P'}^f ({\bf r}_q,{\bf 0})^* \Psi_P^i 
({\bf r}_q, {\bf 0}) d{\bf
r}_q \eeq

\noi where ${\bf r}_Q$ and ${\bf r}_q$ are respectively the positions 
of the active and the spectator quarks.\par

There is a simple argument to find this non-relativistic expression of
the current $J^0 = \overline{Q}\, ' \gamma^0 Q$. Its matrix elements
between one-particle states of given momenta ${\bf p}$ and ${\bf p}\,
'$ in the non-relativistic limit, where ${\bf p}\, '/m_{Q'}$ and ${\bf
p}/m_Q$ are small, are\break\noindent $<{\bf p}\, ', s' |J^0|{\bf p},s > =
\delta_{s',s}$. As readily verified, these matrix elements are
precisely those of the multiplication operator (in configuration space)
by the function $\delta ({\bf r})$. Indeed, one has~:
\beq
<{\bf p}\, ', s'|\delta ({\bf r})|{\bf p}, s> \ = \delta_{s',s} \int 
e^{-i{\bf p}\, '\cdot {\bf r}}\ \delta ({\bf r})\ e^{i{\bf p}\cdot 
{\bf r}} \ d{\bf r} = \delta_{s's} \ .
\eeq

\noi The non-relativistic limit of the current is therefore $J^0 = 
\delta ({\bf r})$.\par

The wave functions, that factorize in center-of-mass and
internal wave functions, are given by
\bea \label{36e} &&\Psi_P^i ({\bf r}_q, {\bf r}_Q) = \exp \left [ i\left ( m_q
{\bf r}_q + m_Q {\bf r}_Q\right ) \cdot {\bf v}\right ] \psi^i \left ({\bf r}_q
- {\bf r}_Q \right ) \nn \\
&&\Psi_{P'}^f ({\bf r}_q, {\bf r}_{Q'}) = \exp \left [ i\left ( m_q
{\bf r}_q + m_{Q'} {\bf r}_{Q'}\right ) \cdot {\bf v'}\right ] \psi^f 
\left ({\bf r}_q
- {\bf r}_{Q'} \right )
\eea

\noi where
\beq \label{37e} {\bf v} = {{\bf P} \over m_q + m_Q} \qquad {\bf v'} =
{{\bf P'} \over m_q + m_{Q'}} \eeq

\noi are the non-relativistic velocities and one gets for the form 
factor (\ref{35e}) the following expression, exhibiting Galilean
invariance, since the NR form factor $F({\bf P}, {\bf P'})$ is a 
function of the variable $({\bf v} - {\bf v'})^2$~:
\beq \label{38e} F({\bf P'} , {\bf P}) = \langle \psi^f \left  |\exp
\left [ i m_q ({\bf v} - {\bf v'})\cdot {\bf r}_q\right ]\right
  | \psi^i\rangle  \equiv f(({\bf v} - {\bf v}{\, '})^2) \ . \eeq

Up to now, equations (\ref{35e})-(\ref{38e}) are valid for the
current $J^0(0) = \overline{Q}\, '(0) \gamma^0 Q(0)$ and for any 
value of the masses
$m_Q$, $m_{Q'}$, $m_q$. If we now assume the hierarchy
\beq
1/R \ll m_q \ll m_{Q}, m_{Q'}
\eeq

\noi we will be in the situation of a heavy-heavy $Q \to Q'$ quark
transition with a light spectator, non-relativistic quark $q$. The
first condition $1/R \ll m_q$, where $R$ is the radius of the bound
state, ensures that the quark $q$ is non-relativistic, while from the
second condition $m_q \ll m_{Q}, m_{Q'}$, the quark $q$ is
light relatively to the active quarks $Q$ and $Q'$. This latter
condition implies a heavy quark symmetry $SU(2N_h)$ where $N_h$ is the
number quarks that are heavy relatively to the quark $q$. Therefore,
these conditions imply Isgur-Wise scaling, with all form factors being
given by the universal form factor $f(({\bf v} - {\bf v'})^2)$
(\ref{38e}). Notably, the flavor independence is due to the fact that 
the internal
wave function, $\psi_f$ and $\psi_i$ become independent of the heavy
quark mass.\par

Expanding the form factor $f(({\bf v} - {\bf v'})^2)$ in powers of 
$({\bf v} - {\bf v'})^2$ we can write then
\beq
f(({\bf v} - {\bf v'})^2) = \sum_n {1 \over n!} f^{(n)} (0) ({\bf v} 
- {\bf v'})^{2n} = \sum_n (-1)^n {1 \over (2n)!} (m_q)^{2n} \ 
<0|z^{2n}|0> \ ({\bf v} - {\bf v'})^{2n}
\eeq

\noi and we obtain the relations between the derivatives of the form 
factor and the moments~:
\beq
f^{(n)}(0) = (-1)^n {n! \over (2n)!} (m_q)^{2n} \ <0|z^{2n}|0>
\eeq

\noi or, by spherical symmetry,
\beq
f^{(n)}(0) = (-1)^n {n! \over (2n+1)} (m_q)^{2n} \mu_{2n} \eeq

\noi with
\beq
\mu_{2n} = <0|r^{2n}|0> \ .
\eeq

\subsection{Relation between the universal NR form factor and the NR 
limit of the Isgur-Wise function.}
\hspace*{\parindent}
Thus, the NR form factor $F({\bf P'}, {\bf P})$ is a
function of the variable $({\bf v} - {\bf v'})^2$, while the
relativistic Isgur-Wise function $\xi (w)$ depends on $w$, where
\beq
\label{13c}
w = v \cdot v' \qquad \hbox{\rm with} \qquad v = {P \over M}\ , \quad 
v' = {P' \over M'}
\eeq

\noi and $(M,P)$, $(M',P')$ are the masses and four-momenta of the 
initial and final mesons.\par

There are several facts that ask for care in the identification between
the NR form factor $f(({\bf v} - {\bf v'})^2)$ and the Isgur-Wise
function $\xi (w)$ in its non-relativistic limit $\xi_{NR}(w)$.\par

Let us leave aside for the moment the fact that the velocities (\ref{13c})
differ from their non-relativistic limits (\ref{37e}) by the binding energy
that is neglected in the latter.\par

Of course, $f(({\bf v} - {\bf v}{\, '})^2)$ and $\xi_{NR}(w)$ cannot 
be generally identified,
since $w$ is not a function of $({\bf v} - {\bf v'})^2$~:
\bea \label{39e} w &=& v \cdot v' = \sqrt{1 + {\bf v'}^2} \ \sqrt{1 +
{\bf v}^2} - {\bf v}\cdot {\bf v'} \nn \\ &=& 1 + {1 \over 2} \left [
({\bf v} - {\bf v'})^2 - \left ( \sqrt{1 + {\bf v}^2} - \sqrt{1 + {\bf
v'}^2}\right )^2 \right ] \ . \eea

To relate $w$ and $({\bf v} - {\bf v'})^2$ one needs to choose a frame.
The natural frame is the rest frame of the initial particle, i. e. ${\bf
v} = 0$.  One has, in this frame~:

\beq \label{40e} w = \left [ 1 + ({\bf v} - {\bf v'})^2 \right]^{1/2} \
. \eeq

\noi The relation between $w$ and $({\bf v} - {\bf v'})^2 $ being
non-linear, the relations between the derivatives relatively to $w$ and
to $({\bf v} - {\bf v'})^2 $ are complicated. As we show in Appendix 
B, the derivative of order $n$ relatively to $w$
depends on the derivatives of order $m \leq n$ relatively to $({\bf v}
- {\bf v'})^2 $ and conversely. \par

Another frame, the equal velocity frame (EVF) where the velocities are
equal and opposite ${\bf v'} = - {\bf v}$, gives from (\ref{39e})~:

\beq \label{41e} w = 1 + {1 \over 2} ({\bf v} - {\bf v'})^2 \ . \eeq

\noi Thus, in this frame, the relation between $w$ and $({\bf v} - {\bf
v'})^2$ is {\it linear}, and the $n$-th derivative relatively to $w$ is
proportional to the $n$-th derivative relatively to $({\bf v} - {\bf
v'})^2 $. \par

In the EVF, we obtain
\bea
\label{42e}
&&\sum_n {1 \over n!} \ \xi_{NR}^{(n)}(1)\ (w-1)^n \nn \\
&&= \xi_{NR}(w) = \langle \psi \left | \exp \left [ i m_q ({\bf v} - 
{\bf v'})\cdot {\bf r}_q \right ] \right | \psi \rangle\nn \\
&&= \sum_{n=0}^{\infty} (-1)^n \ {1 \over (2n)!} \ \left ( 
m_q\right)^{2n} \ <0|z^{2n}|0>\ ({\bf v} - {\bf v'})^{2n} \nn \\
&&= \sum_{n=0}^{\infty} (-1)^n \ 2^n\ {1 \over (2n)!} \ \left ( 
m_q\right)^{2n} \ <0|z^{2n}|0> \ (w-1)^n \ .
\eea

\noi Therefore, in this frame one gets the relation between the 
derivatives of $\xi_{NR}(w)$ and the moments~:

\beq \label{43e}
\xi_{NR}^{(n)}(1) = (-1)^n \ 2^n \ {n! \over (2n+1)!} \ \left ( 
m_q\right)^{2n} \ <0|r^{2n}|0>\ .
\eeq

\noi This relation is exact in this frame, and coincides with the
leading term in the NR limit in all collinear frames, as the rest 
frame (see appendix B).\par

Therefore, in a NR expansion for the light quark, one can only claim to
obtain frame-independent results for the derivatives of $\xi_{NR}(w)$
in the leading NR order for the moments. From now on we will then rely
only on the relation (\ref{43e}). \par

However, ${\bf v}$, ${\bf v'}$ in relation (\ref{42e}) are not identical
to their non-relativistic limits (\ref{37e}). This fact does not
invalidate the relations given above, since $w$ in the EVF (\ref{41e})
differs from its NR expression in terms of NR velocities by terms of
order $\varepsilon/m_Q$ where $\varepsilon$ is the binding energy. This
can be summarized by expanding $w$ in terms of the NR velocities 
(\ref{37e}) and binding energies~:
\beq
w = 1 + {1 \over 2} \left ( {{\bf P} \over m_q + m_Q} - {{\bf P}\, ' 
\over m_q + m_{Q'}}\right )^2 + \hbox{\rm Subleading terms in 
velocities} + O\left ( \varepsilon /m_Q\right )\  .
\eeq

\noi Therefore, since these differences are subleading we can make 
the identification
\beq
\xi_{NR}(w_{NR}) = f(({\bf v} - {\bf v'})^2)
\eeq

\noi where $w_{NR}$ is the leading term of $w$ given by the preceding
expansion, and relation (\ref{43e}) holds indeed for the leading
terms.\par

 From eq. (\ref{43e}) one may be surprised that the NR expansion leads
to a result for $\xi^{(n)}(1)$ increasing with $m_qR \sim (v/c)^{-1}$,
seemingly at odds with the notion of a NR expansion. In fact, the two
last formulas in expression (\ref{42e}) show that the form factor is
expanded in a series of powers $(m_q)^{2n} <0|z^{2n}|0> [({\bf v} - 
{\bf v'})/c]^{2n} \sim (m_q R)^{2n}[({\bf v} - {\bf
v'})/c]^{2n}$ where $R$ is the bound state radius $R \sim
(p_{int})^{-1}$, $p_{int}$ being the {\it internal} quark momentum. On
the other hand ${\bf v} - {\bf v'}$ is the {\it external}
center-of-mass velocity transfer. Therefore, since $m_qR \sim
m_q/p_{int}=[(v/c)_{int}]^{-1}$, the form factor is expanded in powers
of the type $[(v/c)_{int}]^{-2n} [(v/c)_{ext}]^{2n}$. The negative
order in $[(v/c)_{int}]^{-2n}$ is therefore compensated by the
corresponding positive powers of $[(v/c)_{ext}]^{2n}$. The 
derivatives $\xi_{NR}^{(n)}(1)$ in equation
(\ref{43e}) are given by the inverse powers $[(v/c)_{int}]^{-2n}$, that
multiply in (\ref{42e}) $[({\bf v} - {\bf v'})/c]^{2n} = 2^n(w-1)^n$. In a
general frame, the coefficient of $({\bf v} - {\bf v'})^{2n}$ or
$(w-1)^n$ will contain an expansion in powers of
$[(v/c)_{int}]^{-2n+m}$ $(m \geq 0)$. However, the subleading terms $(m
> 0)$ are frame-dependent, as shown in Appendix B.

\section{Constraints on the moments.} \hspace*{\parindent}
Let us define the moments

\beq \label{44e} \mu_n = \ <0|r^n|0> \eeq

\noi and consider the even moments $\mu_{2n}$, related to
$<0|z^{2n}|0>$ from rotational invariance

\beq \label{45e} \mu_{2n} = (2n+1) \ <0|z^{2n}|0> \ . \eeq

\noi We will now formulate {\it necessary} constraints on the 
$\mu_{2n}$ resulting from the fact that they
are indeed moments, i.e. that there exists a function $\varphi (r)$
such that

\beq \label{46e} \mu_{2n} = \int_0^{\infty} r^{2n}|\varphi (r)|^2 dr \
. \eeq

\noi It turns out that these conditions are {\it sufficient}, but this is
only proved in Appendix C, implying that {\it the constraints are
  optimal}. \par

(a) A necessary condition is that for any non-zero polynomial $P$

\beq \label{47e} P(r^2) = \sum_{i=0}^n a_ir^{2i} \geq 0 \quad (r \geq
0) \quad \Rightarrow \int_0^{\infty} P(r^2) |\varphi (r)|^2 dr = 
\sum_{i=0}^n a_i \mu_{2i} > 0\ . \eeq

 From this condition, that is not very explicit, one deduces the
following conditions (b), (c) and (d), that are equivalent. Condition
(d) is explicit.\\

(b) For any $n \geq 0$ and non-vanishing $a_0, \cdots a_n$ one has

\beq \label{48e} \sum_{i,j=0}^n (a_i)^* a_j\ \mu_{2i+2j} > 0 \quad
\hbox{and} \quad \sum_{i,j=0}^n (a_i)^* a_j\ \mu_{2i+2j+2} > 0 \ . \eeq

\noi One demonstrates (b) from (a) by considering the polynomials
$P(r^2) = |\sum\limits_{i=0}^n a_i r^{2i}|^2$ and $P(r^2) = r^2
|\sum\limits_{i=0}^n a_i r^{2i}|^2$. Conversely, (a) results from
(b). \\

(c) For any $n \geq 0$, the matrices $(\mu_{2i+2j})_{0 \leq i, j \leq
n}$ and $(\mu_{2i+2j+2})_{0 \leq i, j \leq n}$ are positive definite.
\par

This condition is just a rephrasing of condition (b).\\

(d) For any $n \geq 0$, one has
\bea
\label{49e}
&&\det \left [ \left ( \mu_{2i+2j}\right )_{0\leq i,j \leq n}\right ] > 0 \\
&&\det \left [ \left ( \mu_{2i+2j+2}\right )_{0\leq i,j \leq n}\right ] > 0 \ .
\label{50e}
\eea

To obtain (d) from (c) it is enough to note that a positive definite
matrix has strictly positive eigenvalues, and that the determinant is
the product of its eigenvalues. \par

Let us first write the determinants (\ref{49e}) and (\ref{50e}) for the
lower values of $n$, namely
\bea
\label{51e}
&&\mu_2 > 0\\
&&\nn \\
\label{52e} &&\det \pmatrix{1 &\mu_2 \cr \mu_2 &\mu_4} > 0\\
&&\nn \\
\label{53e} &&\det \pmatrix{\mu_2 &\mu_4 \cr \mu_4 &\mu_6} > 0\\
&&\nn \\
\label{54e} &&\det \pmatrix{1 &\mu_2 &\mu_4 \cr \mu_2 &\mu_4 &\mu_6\cr
\mu_4 &\mu_6 &\mu_8} > 0\\
&&\nn \\
\label{55e} &&\det \pmatrix{\mu_2 &\mu_4
&\mu_6\cr \mu_4 &\mu_6 &\mu_8\cr \mu_6 &\mu_8 &\mu_{10}} > 0\\
&&\nn \\
\label{56e} &&\det \pmatrix{1 &\mu_2 &\mu_4 &\mu_6\cr \mu_2 &\mu_4
&\mu_6 &\mu_8\cr \mu_4 &\mu_6 &\mu_8 &\mu_{10} \cr \mu_6 &\mu_8
&\mu_{10} &\mu_{12}} > 0\\ && \nn \\ &&\cdots \nn \eea

\noi where (\ref{52e}), (\ref{54e}) and (\ref{56e}) belong to the class
of positivity conditions (\ref{49e}), and (\ref{51e}), (\ref{53e}) 
and (\ref{55e})
to the class (\ref{50e}). \par

 From (\ref{52e}) and (\ref{53e}) we find, respectively \bea \label{57e}
&&\mu_4 > \mu_2^2 \\ &&\nn \\ &&\mu_6 > {\mu_4^2 \over \mu_2} \ . \label{58e}
\eea

\noi To get the constraint on $\mu_8$ from (\ref{54e}) in terms of 
positive definite quantities,
we make use of the following identities,

\beq \label{59e} \mu_4 \det \pmatrix{1 &\mu_2 &\mu_4\cr \mu_2 &\mu_4
&\mu_6\cr \mu_4 &\mu_6 &\mu_{8}} = \det \pmatrix{1 &\mu_2 \cr \mu_2
&\mu_4} \det \pmatrix{\mu_4 &\mu_6 \cr \mu_6 &\mu_8} - \left [ \det
\pmatrix{\mu_2 &\mu_4 \cr \mu_4 &\mu_6}\right ]^2 \eeq

$$\det \pmatrix{1 &\mu_2 &\mu_4\cr \mu_2 &\mu_4
&\mu_6\cr \mu_4 &\mu_6 &\mu_{8}} = $$

\beq \label{60e} \det \pmatrix{\mu_4 &\mu_2 &\mu_6\cr \mu_2 &1
&\mu_4\cr \mu_6 &\mu_4 &\mu_{8}} = \det \pmatrix{\mu_4 &\mu_2 \cr \mu_2
&1} \det \pmatrix{1 &\mu_4 \cr \mu_4 &\mu_8} - \left [ \det
\pmatrix{\mu_2 &1 \cr \mu_6 &\mu_4}\right ]^2 \eeq

\noi that follows from the general identity among determinants of the
Appendix A. We find~:

\beq \label{61e} \mu_8 > {\mu_6^2 - 2 \mu_2 \mu_4 \mu_6 + \mu_4^3
\over \mu_4 - \mu_2^2} = {\left (\mu_2 \mu_6 - \mu_4^2\right )^2 \over
\mu_4 \left (\mu_4 - \mu_2^2\right )} + {\mu_6^2 \over \mu_4} = {\left
(\mu_6 - \mu_2 \mu_4\right )^2 \over \mu_4 - \mu_2^2} + \mu_4^2 \eeq

\noi where the first equality follows from (\ref{54e}) and (\ref{59e})
and the second from (\ref{54e}) and (\ref{60e}). \par

To proceed in the same way with the 10-th moment, we make use of the
inequality (\ref{55e}) and the relations among determinants (A.8)-(A.10).

The inequality (\ref{55e}), together with (A.8)-(A.10)
yields

\beq \label{65e} \mu_{10} > {\mu_2\mu_8^2 - 2 \mu_4 \mu_6 \mu_8 +
\mu_6^3 \over \mu_2 \mu_6- \mu_4^2} = {\left (\mu_4 \mu_8 -
\mu_6^2\right )^2 \over \mu_6 \left (\mu_2\mu_6 - \mu_4^2\right )} +
{\mu_8^2 \over \mu_6} = {\left (\mu_2\mu_8 - \mu_4 \mu_6\right )^2
\over \mu_2\left ( \mu_2 \mu_6 - \mu_4^2\right )} + {\mu_6^2 \over
\mu_2}\ .\eeq

\vskip 3 truemm

\noi Things become more complicated for higher moments, but the method
proceeds in the same way.

\section{Bounds on the derivatives.} \hspace*{\parindent}
Let us summarize the inequalities among the moments deduced in the
previous section. We adopt in (\ref{61e}) and (\ref{65e}) the
expressions given by the last equalities in the r.h.s. This will be
instructive, as it will become clear below. We have obtained
\bea \label{66e} &&\mu_2 > 0 \\
\label{67e} &&\mu_4 > \mu_2^2 \\
\label{68e} &&\mu_6 > {\mu_4^2 \over \mu_2} \\
\label{69e} &&\mu_8 >
{\left ( \mu_6 - \mu_2 \mu_4 \right )^2 \over \mu_4 - \mu_2^2} +
\mu_4^2 \\
&&\mu_{10} > {\left ( \mu_2\mu_8 - \mu_4 \mu_6 \right )^2
\over \mu_2 \left ( \mu_2\mu_6 - \mu_4^2\right ) } +  {\mu_6^2 \over
\mu_2}\ . \label{70e} \eea

 From the {\it frame-independent} relation between moments and
derivatives obtained from (\ref{43e})-(\ref{45e}) (cf. Appendices B and D)~:

\beq \label{71e} \xi_{NR}^{(n)}(1) = (-1)^n \ 2^n {n! \over (2n+1)!} \
\left ( m_q\right)^{2n} \ \mu_{2n}\eeq

\noi and from (\ref{66e})-(\ref{70e}), we obtain
respectively the following inequalities among the derivatives~:
\bea \label{72e} &&- \xi_{NR}^{(1)}(1) > 0 \\ &&\nn \\ \label{73e} &&
\xi_{NR}^{(2)}(1) > {3 \over 5} \ \left [\xi_{NR}^{(1)}(1) \right ]^2
\\ &&\nn \\ &&- \xi_{NR}^{(3)}(1) > - {5 \over 7} \ {\left [\xi_{NR}^{(2)}(1)
\right ]^2 \over \xi_{NR}^{(1)}(1)} \label{74e}
\eea

\beq
\label{75e}
\xi_{NR}^{(4)}(1) >  {7 \over 9}  {\left [ - \xi_{NR}^{(3)}(1) + {3
\over 7}\ \xi_{NR}^{(1)}(1) \ \xi_{NR}^{(2)}(1) \right ]^2 \over
\xi_{NR}^{(2)}(1) - {3 \over 5} \ \left [ \xi_{NR}^{(1)}(1)\right ]^2}
+ {5 \over 21} \left [ \xi_{NR}^{(2)}(1)\right ]^2\eeq

\beq
\label{76e}
- \xi_{NR}^{(5)}(1) >  {9 \over 11} \  {\left [ \xi_{NR}^{(4)}(1) - {5
\over 9}\ {\xi_{NR}^{(2)}(1) \ \xi_{NR}^{(3)}(1) \over
\xi_{NR}^{(1)}(1)} \right ]^2 \over - \xi_{NR}^{(3)}(1)  + {5 \over 7} \
{\left [ \xi_{NR}^{(2)}(1)\right ]^2 \over  - \xi_{NR}^{(1)}(1)} }
- {35 \over 99} \ {\left [ \xi_{NR}^{(3)}(1)\right ]^2 \over
\xi_{NR}^{(1)}(1)} \  .
\eeq

Importantly, we observe that the l.h.s. and the r.h.s. of all the
inequalities (\ref{72e})-(\ref{76e}) scale in the same way in the
parameter $R^2m_q^2$, where $m_q$ is the light quark mass and $R$ the
bound state radius, since the derivatives $\xi_{NR}^{(n)}(1)$, from
(\ref{43e}), scale like $(R^2m_q^2)^n$. The inequalities (\ref{72e}) and
(\ref{73e}) are the non-relativistic limit of the bounds of the true IW
function (\ref{2e}) and (\ref{4e}).\par

Comparing with the results of the method used in Appendix D, we have 
found stronger results
for all the derivatives.\par

For the even derivatives, we find for $2n >
2$ a new term that strengthens the lower bound.
This can be seen from the bounds (\ref{75e}) and (D.9) for
$\xi_{NR}^{(4)}(1)$. We have found a new term besides the term
proportional to $[\xi_{NR}^{(2)}(1)]^2$. However, for the curvature
$\xi_{NR}^{(2)}(1)$, relevant for the non-relativistic limit of the
curvature of the true IW function $\xi (w)$, we
find the same bound (\ref{4e}) as with the former simpler method of Appendix D.
\par

As for the odd derivatives, the trivial bound (D.10) has been
changed in a very substantial way, since we find the lower
bounds (\ref{74e}) and (\ref{76e}). \par

Finally, let us emphasize that the lower bounds (\ref{72e})-(\ref{76e})
are optimal. The optimality of these bounds is demonstrated in the
Appendix C.

\section{Some illustrations.} \hspace*{\parindent}
For the sake of a simple illustration of the bounds, let us consider
the harmonic oscillator in the equal velocity frame~:

\beq \label{77e} \xi_{NR}^{h.o.}(w) = \exp \left [ - (w-1) m_q^2R^2
\right ] \eeq

\noi where the bound state radius $R$ is normalized in a convenient way
to have this simple expression. The $n$-th derivative reads

\beq \label{78e} \xi_{NR}^{(n)}(1) = (-1)^n \left ( m_q^2 R^2 \right
)^n \ . \eeq

\noi Then, the bound (\ref{72e}) reads simply $m_q^2R^2 > 0$ and
(\ref{73e})-(\ref{76e}) will become for the $n$-th derivative
\bea
\label{79e}
&n = 2 &\qquad 1 > {3 \over 5} \nn \\
&n = 3 &\qquad 1 > {5 \over 7} \nn \\
&n = 4 &\qquad 1 > {55 \over 63} \nn \\
&n = 5 &\qquad 1 > {91 \over 99} \ .
\eea

Interestingly, we find that the bounds become better and better as we
consider higher derivatives. For the 5-th derivative the bound is already very
strict. \par

However, it is not granted that these features will remain for
more realistic potentials. Therefore, it can be useful to examine 
another simple potential,
although not confining, namely the Coulomb potential. In this case we
have a dipole form factor

\beq \label{80e}
\xi_{NR}^{Coulomb}(w) = {1 \over \left [ 1 + (w-1)m_q^2R^2\right ]^2} \ .
\eeq

\noi The derivatives read

\beq
\label{81e}
\xi_{NR}^{(n)}(1) = (n+1)!\ (-1)^n\  \left ( m_q^2R^2\right )^n
\eeq

\noi and the inequalities (\ref{73e})-(\ref{76e}) give, respectively
\bea
\label{82e}
&n = 2 &\qquad 1 > {2 \over 5} \nn \\
&n = 3 &\qquad 1 > {15 \over 28} \nn \\
&n = 4 &\qquad 1 > {263 \over 378} \nn \\
&n = 5 &\qquad 1 > {3626 \over 4719} \ .
\eea

These inequalities are somewhat less strict than in the harmonic
oscillator case but here also they improve for higher derivatives.
\par

We can expect that in the case of a realistic phenomenological
$Q\bar{q}$ potential, with a confining and a short distance parts, the
situation will be in between the harmonic oscillator and the Coulomb
potentials.

\section{Generalization of the method to QCD.} \hspace*{\parindent}
We have obtained lower bounds on the derivatives
at zero recoil of the non-relativistic Isgur-Wise function
$\xi_{NR}(w)$, i.e. the IW function with a NR light quark. Our main
motivation has been to find the leading term in a NR expansion of the
derivatives at zero recoil of the true IW function $\xi (w)$ that
should be obtained in the heavy quark limit of QCD. The parameter in
this expansion is $(v^2/c^2)_{int}$ or, equivalently $1/R^2m_q^2$, where $m_q$
is the light quark mass and $R$ is the bound state radius. In previous
work \cite{3r} we did obtain in the heavy quark limit of QCD such an expansion
for the slope and the curvature, inequalities (\ref{2e}) and (\ref{4e}),
\bea \label{83e} &&- \xi^{(1)} (1) > {3 \over 4} \nn \\ &&\xi^{(2)} (1) >
{1 \over 5} \left \{ - 4 \xi^{(1)} (1) + 3 \left [ \xi^{(1)} (1) \right
]^2 \right \} \ . \eea

\noi Since $- \xi^{(1)} (1)$ and $\xi^{(2)} (1)$, scale
respectively like $R^2m_q^2$ and $(R^2 m_q^2)^2$, in the NR limit these
inequalities become respectively (\ref{72e}) and (\ref{73e}). The
inequalities (\ref{83e}) contain terms, specific to QCD in the heavy
quark limit, that are subleading in a NR expansion. \par

Our aim would be, in the long run, to obtain bounds for the $n$-th
derivative of the IW function in the heavy quark limit of QCD that must
contain the subleading terms in a NR expansion. We know that in the
strict NR limit we must recover the bounds (\ref{72e})-(\ref{76e})
obtained in the present paper. \par

To obtain these bounds in QCD we could try the method of \cite{3r} in 
a systematic way, that uses sum rules for the
non-forward amplitude, relating a sum over intermediate states and the
OPE, that depends on three variables $w_i = v_i \cdot v'$, $w_f = v_j
\cdot v'$, $w_{if} = v_i \cdot v_f$, that lie in a
certain domain \cite{1r}. Differentiating the SR relatively to $(w_i,
w_f, w_{if})$ and going to the frontier of the domain one gets
relations that allow to obtain (\ref{83e}) \cite{3r}. This method can
be pursued further and obtain bounds for the higher derivatives
\cite{6r}. However, the obtained bounds, in their NR limit, are 
weaker than (\ref{74e})-(\ref{76e}).\par

We have developed here a more powerful method, based on the
positivity of matrices of moments of the ground state wave function,
that allows to go further for the derivatives $n > 2$ in the NR 
limit. To generalize the present method to QCD in the heavy quark 
limit one
should investigate whether the derivatives $\xi^{(n)}(1)$ can be expressed
in terms of positive definite quantities that are true moments as in 
the non-relativistic expression (\ref{46e}).
Then, one could draw the consequences that follow from
the positivity of the relevant matrices. A step in this direction is
the conjecture that, at least in the meson case \cite{1r}, all SR in 
the heavy quark limit of QCD
are satisfied in the Bakamjian-Thomas (BT) class of relativistic quark
models \cite{7r}. We have realized this in practice for the lower
derivatives $\xi^{(n)}(1)$, $n = 1,2,3$. These models are 
relativistic for the states and also for
the current matrix elements in the heavy quark limit, exhibiting
Isgur-Wise scaling. One can hope to start from the NR quark model and
go to BT models, and from those to the heavy quark limit QCD. \par

Another, more direct way to proceed to the heavy quark limit of QCD is
to start from the sum rules obtained in \cite{1r}-\cite{3r}, 
realizing that one can obtain the NR bounds of the present paper from
the equivalent sum rules of the non-relativistic limit. Indeed, in 
the NR limit we have a SR of the form

\beq \label{(2)} \sum_{n''} f_{n,n''}({\bf k})f_{n'',n'}({\bf k'}) =
f_{n,n'}({\bf k} + {\bf k'}) \ . \eeq

\noi that follows, very simply, from

\beq \label{(1)} f_{n,n'}({\bf k}) = \ < n|e^{i{\bf k}\cdot {\bf r}}|n'> \ .
\eeq

\noi In QCD in the heavy quark limit we have sum rules of the same form
(\ref{(2)}), but {\it without} the explicit expression (\ref{(1)}). However, to
derive the inequalities of this paper, (\ref{(2)}) is sufficient. 
Indeed, from (\ref{(2)}) one
gets

\beq \label{(4)} \sum_{i,j} c_ic_j^*f_{0,0}\left ( {\bf k}_i - {\bf
k}_j\right ) = \sum_n \sum_{i,j} c_ic_j^* f_{0,n}({\bf k}_i) 
f_{n,0}(- {\bf k}_j) = \sum_n | \sum_i c_i f_{0,n}({\bf k}_i)|^2 \geq 
0 \ . \eeq

\noi From this relation one can infer, for any function $\varphi({\bf k})$,

\beq \label{(5)} \int d{\bf k} \ d{\bf k'} \ \varphi ({\bf k'})^*\
f_{0,0}({\bf k} - {\bf k'})\ \varphi ({\bf k}) \geq 0 \eeq

\noi and therefore, for the Fourier transform

\beq \label{(6)} \int d{\bf r} \ | \widetilde{\varphi}({\bf r})|^2 \
\widetilde{f}_{0,0}({\bf r}) \geq 0 \eeq

\noi and hence

\beq \label{(7)} \widetilde{f}_{0,0}({\bf r}) \geq 0 \ . \eeq

\noi These are the conditions that we need to obtain constraints on the
moments and hence bounds on the derivatives of the form factor
$f_{0,0}({\bf k} - {\bf k'})$, because writing the form factor in 
terms of its Fourier transform

\beq
\label{58t}
f_{0,0}({\bf k} - {\bf k'}) = \int d{\bf r}\ \widetilde{f}_{00}({\bf 
r}) \ e^{i({\bf k} - {\bf k'})\cdot {\bf r}}
\eeq

\noi and taking into account that $f_{0,0}({\bf k} - {\bf k'})$ must be
an even function, we obtain (Oz is defined along the momentum transfer
${\bf k} - {\bf k'}$)~:

\beq
\label{59t}
f_{0,0}({\bf k} - {\bf k'}) = \sum_n (-1)^n \ {1 \over (2n)!} \ ({\bf 
k} - {\bf k'})^{2n} \ {\mu_{2n} \over 2n+1}
\eeq

\noi i.e. an expansion of the form factor in terms of moments of the 
form (\ref{42e}), with the identification

\beq
\label{60t}
\mu_{2n} = \ <0|r^{2n}|0> \ = \int d{\bf r} \ 
\widetilde{f}_{0,0}({\bf r})\ r^{2n}
\eeq
\vskip 3 truemm

\noi since the positivity condition $\widetilde{f}_{0,0}({\bf r}) \geq
0$ (\ref{(7)}) holds and, from (\ref{58t})~:

\beq
\int
\widetilde{f}_{0,0}({\bf r}) d{\bf r} = 1\ .
\eeq

We do recover essentially the previous results (\ref{66e})-(\ref{70e}) {\it
using only the sum rules} (\ref{(2)}). If $\widetilde{f}_{0,0}({\bf r})$
is a function, it can be seen as the square of a wave function, and all
the {\it strict} inequalities of the NR type (\ref{66e})-(\ref{70e}) would
follow. However, this is not implied by the SR, and weaker results
could follow, namely the inequalities may not be strict.
For example, one could have a distribution like

\beq
\label{61x}
\widetilde{f}_{0,0}({\bf r}) = {1 \over 4 \pi r_0^2} \delta (|{\bf r}| - r_0)
\eeq

\noi which is not the square of a wave function, that would imply

\beq
\label{62x}
\mu_{2n} = r_0^{2n}
\eeq

\noi and, for example, the strict inequality (\ref{67e}) would become an
equality. For example, in the true QCD case, the lower bound 
(\ref{2e}) for $\rho^2$ could
become an equality. By the way, this would correspond to the so-called
BPS approximation \cite{10r}.

Our strategy will then be to start from the SR in the heavy quark limit
of QCD that are equivalent to the NR ones (\ref{(2)}), and proceed along
the same lines. We can presume that the method will give the optimal
bounds for the derivatives of the true Isgur-Wise function $\xi (w)$.

\section{Conclusion.}
\hspace*{\parindent}
To conclude, we have obtained the best possible general bounds on the 
derivatives of the
Isgur-Wise function $\xi_{NR}(w)$, i.e. considering the light quark as
non-relativistic, in terms of lower derivatives. These bounds must be the
non-relativistic limit of the bounds on the derivatives of the true
Isgur-Wise function $\xi (w)$, and constitute a guideline in the
derivation of the latter. Moreover, we argue that the method developed
here, that exploits the positivity of matrices of moments can be
generalized, starting from SR in the heavy quark limit of QCD, to
obtain the best bounds on all the derivatives of $\xi (w)$.

\newpage
\section*{Appendix A. An identity between determinants.}
\hspace*{\parindent}
In Section 3 we have made use of the identity among determinants
$$\det \left [ (a_{ij})_{1\leq i, j \leq n}\right ] \ \det \left [
(a_{ij})_{2\leq i, j \leq n-1}\right ] = \det \left [ (a_{ij})_{1\leq i,
j \leq n-1}\right ]\ \det \left [ (a_{ij})_{2\leq i, j \leq n}\right ]$$
$$- \det \left [ (a_{ij})_{1\leq i\leq n-1 ,2 \leq  j \leq n}\right ]\
\det \left [ (a_{ij})_{2\leq i\leq n, 1 \leq j \leq n-1}\right ]\eqno({\rm
A.1})$$

\noi or, in a more readable way~:

$$\det \pmatrix{a_{1,1} &a_{1,2} &\cdots &a_{1,n-1} &a_{1,n}\cr a_{2,1}
&a_{2,2} &\cdots &a_{2,n- 1} &a_{2,n}\cr \cdots &\cdots &\cdots &\cdots
&\cdots \cr a_{n-1,1} &a_{n-1,2} &\cdots &a_{n-1,n- 1} &a_{n-1,n}\cr
a_{n,1} &a_{n,2} &\cdots &a_{n,n- 1} &a_{n,n}\cr} \det \pmatrix{a_{2,2}
&\cdots &a_{2,n- 1}\cr \cdots &\cdots &\cdots \cr a_{n- 1,2} &\cdots
&a_{n-1,n-1}\cr}$$ $$=\det \pmatrix{a_{2,2} &\cdots &a_{2,n-1} &a_{2,n}\cr
\cdots &\cdots &\cdots &\cdots \cr a_{n-1,2} &\cdots &a_{n-1,n- 1}
&a_{n-1,n}\cr a_{n,2} &\cdots &a_{n,n- 1} &a_{n,n}\cr} \det
\pmatrix{a_{1,1} &\cdots &a_{1,n-2} &a_{1,n-1}\cr \cdots &\cdots
&\cdots &\cdots \cr a_{n-2,1} &\cdots &a_{n-2,2} &a_{n-2,n-1}\cr
a_{n-1,1} &\cdots &a_{n- 1,2} &a_{n-1,n-1}\cr}$$ $$-\det \pmatrix{a_{1,2}
&\cdots &a_{1,n-1} &a_{1,n}\cr \cdots &\cdots &\cdots &\cdots \cr
a_{n-2,2} &\cdots &a_{n-2,n- 1} &a_{n-2,n}\cr a_{n-1,2} &\cdots
&a_{n-1,n- 1} &a_{n-1,n}\cr} \det \pmatrix{a_{2,1} &\cdots &a_{2,n-2}
&a_{2,n-1}\cr \cdots &\cdots &\cdots &\cdots \cr a_{n-1,1} &\cdots
&a_{n-1,n-2} &a_{n-1,n-1}\cr a_{n,1} &\cdots &a_{n,n-2} &a_{n,n-1}\cr} \ .
\eqno({\rm A.2})$$

\noi To demonstrate this relation, let us introduce the column vectors

$$x_i = \sum_{j=1}^n a_{i,j} \ e_j \eqno({\rm A.3})$$

\noi where the $a_{i,j}$ are the elements of the matrices (A.1) or
(A.2). Multipying (A.2) by $(e_1 \wedge \cdots \wedge e_n) \otimes (e_1
\wedge \cdots \wedge e_n)$, this formula writes

$$\left ( x_1 \wedge x_2 \wedge \cdots \wedge x_{n-1} \wedge x_n \right
) \otimes \left ( e_1 \wedge x_2 \wedge \cdots \wedge x_{n-1} \wedge
e_n \right )$$ $$= \left ( x_1 \wedge x_2 \wedge \cdots \wedge x_{n-1}
\wedge e_n \right ) \otimes \left ( e_1 \wedge x_2 \wedge \cdots \wedge
x_{n-1} \wedge x_n \right )$$ $$- \left ( e_n \wedge x_2 \wedge \cdots
\wedge x_{n-1} \wedge x_n \right )\otimes \left ( x_1 \wedge x_2 \wedge
\cdots \wedge x_{n-1} \wedge e_1 \right )\ .\eqno({\rm A.4})$$

\noi Assuming that the vectors $x_1 , \cdots x_n$ are independent, one
can expand $e_1$ and $e_n$~:

$$e_1 = \alpha_1 x_1 + \cdots + \alpha_n x_n \qquad e_n = \beta_1 x_1 +
\cdots + \beta_n x_n\ .\eqno({\rm A.5})$$

\noi The l.h.s. of (A.4) becomes

$$\left ( x_1 \wedge x_2 \wedge \cdots \wedge x_{n-1} \wedge x_n \right
) \otimes \left ( e_1 \wedge x_2 \wedge \cdots \wedge x_{n-1} \wedge
e_n \right )$$ $$= \left ( x_1 \wedge x_2 \wedge \cdots \wedge x_{n-1}
\wedge x_n \right ) \otimes \left ( \left ( \alpha_1 x_1 + \alpha_n
x_n\right ) \wedge x_2 \wedge \cdots \wedge x_{n-1} \wedge \left (
\beta_1 x_1 + \beta_n x_n \right ) \right ) $$ $$= \left (
\alpha_1 \beta_n - \alpha_n \beta_1 \right ) \left ( x_1 \wedge x_2
\wedge \cdots \wedge x_{n-1} \wedge x_n\right ) \otimes \left ( x_1
\wedge x_2 \wedge \cdots \wedge x_{n-1} \wedge x_n\right ) \eqno({\rm
A.6})$$

\noi while the terms in the r.h.s. become
$$\left ( x_1 \wedge x_2 \wedge \cdots \wedge x_{n-1} \wedge e_n\right
) \otimes \left ( e_1 \wedge x_2 \wedge \cdots \wedge x_{n-1} \wedge
x_n\right )$$ $$=  \alpha_1 \beta_n \left ( x_1 \wedge x_2 \wedge
\cdots \wedge x_{n-1} \wedge x_n\right ) \otimes \left ( x_1 \wedge x_2
\wedge \cdots \wedge x_{n-1} \wedge x_n\right )$$

  $$\left ( e_n \wedge x_2 \wedge \cdots \wedge x_{n-1} \wedge x_n\right
) \otimes \left ( x_1 \wedge x_2 \wedge \cdots \wedge x_{n-1} \wedge
e_1\right )$$ $$=  \alpha_n \beta_1 \left ( x_1 \wedge x_2 \wedge
\cdots \wedge x_{n-1} \wedge x_n\right ) \otimes \left ( x_1 \wedge x_2
\wedge \cdots \wedge x_{n-1} \wedge x_n\right )\ . \eqno({\rm A.7})$$

\noi The identity is therefore demonstrated if the vectors $x_1, \cdots
x_n$ are independent. \par

If the vectors are dependent one can show that both the l.h.s. and the
r.h.s. vanish identically.\par

In particular, we make use in Section 3 of the following identities~:

$$\det \pmatrix{\mu_4 &\mu_6 \cr \mu_6 &\mu_8} \det
\pmatrix{1 &\mu_2 &\mu_4 &\mu_6\cr \mu_2 &\mu_4 &\mu_6 &\mu_8\cr \mu_4
&\mu_6 &\mu_8 &\mu_{10} \cr \mu_6 &\mu_8 &\mu_{10} &\mu_{12}}$$
$$=
\det \pmatrix{\mu_4 &\mu_6 &\mu_8\cr \mu_6 &\mu_8 &\mu_{10}\cr \mu_8
&\mu_{10} &\mu_{12}}\det \pmatrix{1 &\mu_2 &\mu_4 \cr \mu_2 &\mu_4
&\mu_6\cr \mu_4 &\mu_6 &\mu_8} - \left [ \det \pmatrix{\mu_2 &\mu_4
&\mu_6\cr \mu_4 &\mu_6 &\mu_8\cr \mu_6 &\mu_8 &\mu_{10}}\right ]^2 
\eqno({\rm A.8})$$

$$\det \pmatrix{1 &\mu_4 \cr \mu_4 &\mu_8} \det
\pmatrix{\mu_4 &\mu_2 &\mu_6 &\mu_8\cr \mu_2 &1 &\mu_4 &\mu_6\cr \mu_6
&\mu_4 &\mu_8 &\mu_{10} \cr \mu_8 &\mu_6 &\mu_{10} &\mu_{12}}$$
$$=
\det \pmatrix{\mu_4 &\mu_2 &\mu_6\cr \mu_2 &1 &\mu_{4}\cr \mu_6
&\mu_{4} &\mu_{8}}\det \pmatrix{1 &\mu_4 &\mu_6 \cr \mu_4 &\mu_8
&\mu_{10}\cr \mu_6 &\mu_{10} &\mu_{12}} - \left [ \det \pmatrix{\mu_2
&1 &\mu_4\cr \mu_6 &\mu_4 &\mu_8\cr \mu_8 &\mu_6 &\mu_{10}}\right ]^2
\eqno({\rm A.9})$$

$$\det \pmatrix{\mu_4 &\mu_2 \cr \mu_2 &1} \det
\pmatrix{\mu_8 &\mu_6 &\mu_4 &\mu_{10}\cr \mu_6 &\mu_4 &\mu_2 &\mu_8\cr
\mu_4 &\mu_2 &1 &\mu_{6} \cr \mu_{10} &\mu_8 &\mu_{6} &\mu_{12}}$$
$$= \det \pmatrix{\mu_8 &\mu_6 &\mu_4\cr \mu_6 &\mu_4 &\mu_{2}\cr \mu_4
&\mu_{2} &1}\det \pmatrix{\mu_4 &\mu_2 &\mu_8 \cr \mu_2 &1 &\mu_6\cr
\mu_8 &\mu_6 &\mu_{12}} - \left [ \det \pmatrix{\mu_6 &\mu_4 &\mu_2\cr
\mu_4 &\mu_{12}&1\cr \mu_{10} &\mu_{8} &\mu_{6}}\right ]^2 . 
\eqno({\rm A.10})$$

\vskip 1 truecm
\section*{Appendix B. Frame-dependence of the subleading moments.}
\hspace*{\parindent} Let us begin from the general expression
(\ref{38e}), (\ref{40e}) in the initial hadron rest frame, ${\bf v} = 
0$, without neglecting powers
$<0|z^{2p}|0> (w-1)^p$ ($p < m$). From the relation

$$\left ( {\bf v} - {\bf v}'\right )^2 = w^2 - 1 = \sum_{i=0}^m 2^{m-i} {m
\choose i} (w-1)^{m+i} \eqno({\rm B.1})$$

\noi one obtains the following relation between
derivatives and moments~:

$$\xi_{NR}^{(n)}(1)= \sum_{m=n/2}^n (-1)^m\ 2^{2m-n}\
{n!m! \over (2m)!(n-m)!(2m-n)!} \ \left ( m_q\right )^{2m} \
<0|z^{2m}|0> \ . \eqno({\rm B.2})$$

\noi conversely, to obtain the moments in terms of the derivatives, it
is enough to replace the variable $w$ by

$$y = w^2 - 1 \ .\eqno({\rm B.3})$$

\noi Then
$$\xi_{NR}(w) = \xi_{NR}[(1+y)^{1/2}] \eqno({\rm B.4})$$

\noi can be expanded in
powers of $y = w^2-1$~:
$$\xi_{NR} [(1+y)^{1/2}] = \sum_{n =0}^{\infty} {1
\over n!} \ \xi_{NR}^{(n)}(1) \sum_{m =n}^{\infty} c_{n,m}\ y^m$$
$$= \sum_{m =0}^{\infty} (-1)^m\ {1 \over (2m)!} \ \left ( m_q\right
)^{2m} \ <0|z^{2m}|0> \ y^m\ . \eqno({\rm B.5})$$

\noi From this expression we can read the expression of the moments in
terms of the derivatives

$$<0|z^{2m}|0> = (-1)^m \ (2m)! \left ( m_q\right
)^{-2m} \sum_{n =0}^m {1 \over n!} \ c_{n,m}\ \xi_{NR}^{(n)}(1)   \ .
\eqno({\rm B.6})$$

The coefficients $c_{n,m}$ are defined by

$$ \left [ (1+y)^{1/2}-1 \right ]^n = \sum_{m=n}^{\infty}
c_{n,m}\ y^m \eqno({\rm B.7})$$

\noi and therefore given by

$$ c_{n,m} = \sum_{i=0}^n (-1)^{n-i} {n \choose i} {i/2
\choose m}\ . \eqno({\rm B.8})$$

\noi This sum can be calculated and gives~:

$$ c_{n,m} = (-1)^{m-n } \ 2^{-2m+n} \left [ {2m- n-1
\choose m-n} - {2m-n-1 \choose m-n-1}\right ] \ . \eqno({\rm B.9})$$

\noi Explicitly, one has

$$c_{0,m} = \delta_{m,0}$$
$$c_{n,m} = (-1)^{m-n
} \ 2^{-2m+n}\ n\ {(2m-n-1)! \over m!(m-n)!} \eqno({\rm B.10})$$

\noi where the second relation holds except for $n = m = 0$, since
$c_{0,0} = 1$. \par

Gathering relations (B.6)-(B.10) we obtain the final
relation giving the moments in terms of the derivatives
$$<0|z^{2m}|0> = \left ( m_q\right )^{-2m} \Big [
\delta_{m,0}\ \xi_{NR}(1)$$
$$+ \sum_{n =1}^m (-1)^{n } \
2^{-2m+n}\ {(2m)! (2m-n-1)! \over m!(m-n)!(n-1)!}  \ \xi_{NR}^{(n)}(1)\
. \eqno({\rm B.11})$$

\noi The relations (B.2) and (B.11) are the main results of
this section. \par

Explicitly, one obtains, for the lower derivatives and moments~:
$$\xi_{NR}(1) = \ <0|1|0>$$
$$\xi '_{NR}(1) = -m_q^2\ <0|z^2|0>$$
$$\xi ''_{NR}(1) = {1 \over 3}\ m_q^4 \ <0|z^4|0> - \ m_q^2\ <0|z^2|0>$$
$$\xi_{NR}^{(3)}(1) = - {1 \over 15
}\ m_q^6 \ <0|z^6|0> + \ m_q^4\ <0|z^4|0>$$
$$\xi_{NR}^{(4)}(1) = {1
\over 105 }\ m_q^8 \ <0|z^8|0> - {2 \over 5} \ m_q^6\ <0|z^6|0> +
m_q^4\ <0|z^4|0>$$
$$\xi_{NR}^{(5)}(1) = - {1 \over 945 }\ m_q^{10}
\ <0|z^{10}|0> + {2 \over 21} \ m_q^8\ <0|z^8|0> - m_q^6\ <0|z^6|0>
\eqno({\rm B.12})$$

$$<0|1|0> = \xi_{NR}(1)$$
$$<0|z^2|0> = - m_q^{-2}\ \xi '_{NR}(1)$$
$$<0|z^4|0> = 3 \ m_q^{-4}\left [ \xi ''_{NR}(1) - \xi '_{NR}(1)\right ]$$
$$<0|z^6|0> = 15\  m_q^{-6}\left [ - \xi^{(3)}_{NR}(1) + 3\ \xi 
''_{NR}(1) - 3 \ \xi
'_{NR}(1)\right ]$$
$$<0|z^8|0> = 105 \ m_q^{-8}\left [
\xi^{(4)}_{NR}(1) -6 \ \xi_{NR}^{(3)}(1) + 15\ \xi ''_{NR}(1) - 15\  \xi
'_{NR}(1)\right ]$$
$$<0|z^{10}|0> = 945 \ m_q^{-10}\left [ -
\xi^{(5)}_{NR}(1) +106 \ \xi_{NR}^{(4)}(1) - 45\ \xi_{NR}^{(3)}(1)\right.$$
$$\left. + 105 \ \xi ''_{NR}(1) - 105 \ \xi '_{NR}(1)\right ] \ . 
\eqno({\rm B.13})$$

We observe that in expressions (B.2) and (B.11) there is a
leading term in the non-relativistic expansion and in (B.12) and 
(B.13), the first term
in the expansion is the leading term. \par

The purpose of the present detailed calculation is to point out that
actually the subleading terms are {\it frame-dependent} and that
therefore, one can only get information on the {\it leading term} in
the non-relativistic expansion, the unique term that appears in the 
equal-velocity-frame used in Section 2.\par

Moreover, there is continuity between the rest frame and the
equal-velocity frame, since $w$ can be expressed in terms of $({\bf v}
- {\bf v}')^2$ in all collinear frames

$$\alpha {\bf v} + \beta {\bf v}' = 0 \quad , \qquad \alpha +
\beta = 1 \eqno({\rm B.14})$$

\noi one gets indeed

$$w = \sqrt{1 + (1 - \alpha )^2({\bf v} - {\bf v}')^2} \ \sqrt{1 +
\alpha^2 ({\bf v} - {\bf v}')^2} + \alpha (1 - \alpha ) ({\bf v}
- {\bf v}')^2 \eqno({\rm B.15})$$

\noi with $\alpha = 1$ in the rest frame and $\alpha = {1 \over 2}$ in
the equal-velocity-frame. The first order in $({\bf v} - {\bf v}')^2$ 
yields, independently of $\alpha$,

$$w \cong 1 + {1 \over 2} ({\bf v} - {\bf v}')^2 \eqno({\rm
B.16})$$

\noi giving the leading order relation between derivatives of
$\xi_{NR}(w)$ and moments (\ref{43e}).\par

\vskip 1 truecm
\section*{Appendix C. Optimality of the constraints.} \hspace*{\parindent}
We have seen that the non-relativistic Isgur-Wise function

$$\xi_{NR}(w) = \ <\psi_0|e^{-im_q({\bf v}' - {\bf v})\cdot
{\bf r}}|\psi_0> \qquad \left [ w = {1 \over 2} ({\bf v}' - {\bf v})^2 +
1 \right ] \eqno({\rm C.1})$$

\noi has its derivatives at $w = 1$ related to the moments

$$\mu_{2n} = \ <\psi_0|r^{2n}|\psi_0> \eqno({\rm C.2})$$

\noi by

$$\xi_{NR}^{(n)}(1) = (-1)^n \ {2^n n! \over (2n+1)!}\ m_q^{2n} \
\mu_{2n} \eqno({\rm C.3})$$

\noi and that these moments satisfy the following tower of
constraints~:

$$\det \left ( \mu_{2i+2j}\right )_{0\leq i,j \leq n} > 0 \eqno({\rm
C.4})$$

$$\det \left ( \mu_{2i+2j+2}\right )_{0\leq i,j \leq n} > 0 \eqno({\rm
C.5})$$

\noi which consists, for each $n \geq 0$, of a lower bound on
$\mu_{2n}$ depending on the moments $\mu_{2k}$ for $0 \leq k \leq n-1$.
\par

In this appendix, we show that this cannot in general (for arbitrary
wave function) be improved. Namely that, given $\mu_{2k}$ for $0 \leq k
\leq n-1$ satisfying (C.4) and (C.5), the moment $\mu_{2n}$ can have
{\it any} value larger than this lower bound. \par

To that goal, we forget (C.2) by now, and consider {\it arbitrary}
numbers $\mu_{2n}$ $(n = 0, 1, 2, \cdots )$ satisfying the constraints
(C.4) and (C.5). In this appendix we prove that, for {\it any} $N \geq
0$, there exists a wave function $\psi_0^{(N)}$ such that one has

$$\mu_{2n} = \ < \psi_0^{(N)} |r^{2n}|\psi_0^{(N)}> \qquad \hbox{for all
$0 \geq k \geq N$} \ . \eqno({\rm C.6})$$

\noi We shall not be able here to know if there is a wave function
$\psi_0$ satisfying (C.2) for all $n$, but our more limited result
(C.6) is enough to prove the point.\\

To simplify notations, we introduce

$$x = r^2 \eqno({\rm C.7})$$

\noi as a variable taking
positive values. \\

{\bf 1 -} Introduce, in the vector space of polynomials in $x$, the
linear form defined by the values $\mu_{2k}$ on the monomials $x^k$
(which constitutes an algebraic basis of this vector space). It is
given by

$$<P> \ = \sum_{k=0}^n a_k \ \mu_{2k} \qquad \hbox{for} \quad P(x) =
\sum_{k=0}^n a_k\ x^k \ .\eqno({\rm C.8})$$

\noi As a preliminary and crucial step, we have~:\\

{\bf 2 - } This linear form $<P>$ is strictly positive. Namely, one has

$$P(x) \geq 0 \ \hbox{for all $x \geq 0$, \quad and $P\not = 0
\Rightarrow \ <P> \ > 0$}\ .\eqno({\rm C.9})$$

To prove this, observe first that (C.4) and (C.5) imply that
$(\mu_{2i+2j})_{0 \leq i, j\leq n}$ and $(\mu_{2i+2j+2})_{0 \leq i,
j\leq n}$ are positive definite matrices, or explicitely that one has

$$\sum_{i,j=0}^n a_i\ a_j^* \ \mu_{2i+2j} > 0 \ , \qquad 
\sum_{i,j=0}^n a_i\ a_j^* \ \mu_{2i+2j+2} > 0 \eqno({\rm C.10})$$

\noi for any coefficients $a_0, \cdots , a_n$ not all vanishing, and
using the definition (C.8) of $<P>$, these properties (C.10) translate
into~:

$$<|Q|^2> \ > 0\ , \qquad <x|Q|^2> \ > 0 \eqno({\rm C.11})$$

\noi for any non-vanishing polynomial $Q$.\par

Then, any $P$ satisfying $P(x) \geq 0$ for all $x \geq 0$ is a linear
combination with positive coefficients of polynomials of the form
$|Q|^2$ or $x|Q|^2$. Indeed, considering the roots of $P$, we have

$$P = \prod\nolimits_i (x + c_i) \prod\nolimits_j (x - c'_j)^2 
\prod\nolimits_k |x - z_k|^2 \eqno({\rm C.12})$$

\noi with $c_i \geq 0$, $c'_j > 0$, Im $z_k \not= 0$, since complex
roots $z_k$ occur in conjugate pairs, strictly positive roots $c'_j$
occur in event multiplicity (else a change of sign at $x = c'_j$), and
negative roots $-c_i$ are arbitrary.\\

{\bf 3 - } Next introduce a scalar product in the vector space of 
polynomials by

$$<P|P'> \ = \ <P^*P'> \ . \eqno({\rm C.13})$$

\noi The scalar product properties are easily verified. Notably, the
important fact that $<P|P> = 0$ implies $P = 0$ results from (C.4).
\par

We may then consider the orthogonal polynomials $p_0, p_1, p_2, \cdots$
with respect to this scalar product. The theory of orthogonal
polynomials is classsical matter \cite{8r}. They are usually
considered with respect to a scalar product defined by a weighted
integral, but their properties extends easily to the more general case
needed here, where we do not know a priori if the scalar product (C.13)
can be given by an integral.\\

The polynomial $p_n$ has degree $n$, and we have~:

$$<p_n|p_{n'}> = d_n^2\ \delta_{n,n'}\ . \eqno({\rm C.14})$$

\noi It will be convenient for us to fix $p_n$ by taking the
coefficient of $x^n$ to be 1. These polynomials can be computed
recursively by the orthogonalisation Schmidt process~:

$$p_n = x^n - \sum_{k=0}^{n-1} {<p_k|x^n> \over <p_k|p_k>}\ p_k
\eqno({\rm C.15})$$

\noi where the automatic fact that $<p_k|p_k > \not= 0$ is essential.
Also, since any polynomial of degree $\leq n$ is a linear combination
of $p_0, \cdots, p_n$, we have the property~:

$$<P|p_n> \ = 0 \qquad \hbox{for any $P$ of degree $ < n$} \eqno({\rm
C.16})$$

\noi which is of constant use in the following. Taking $P = 1, x,
\cdots, x^{n-1}$, (C.16) gives a system of $n$ linear equations for the
$n+1$ coefficients of $p_n$, which, according to (C.4), can be solved
uniquely up to a constant, and then Cramer's formulae give an
explicit expression for $p_n$. \\

We are actually interested by the zeros of $p_n$. \\

{\bf 4 -} All the roots of $p_n$ are simple and strictly positive. \\

In fact, let $x_1, \cdots, x_m$ be the strictly positive roots of $p_n$ of
{\it odd} multiplicity. We have only to show that $m = n$. If $m < n$,
according to (C.16), we have

$$<(x-x_1) \cdots (x - x_m) p_n > \ = \ < (x-x_1) \cdots (x- x_m)|p_n > \ =
0\ . \eqno({\rm C.17})$$

\noi However, the polynomial $(x - x_1) \cdots (x - x_m)p_n$ has a
constant sign for $x \geq 0$, and does not vanish identically.
Therefore, according to (C.9), (C.17) and hence $m < n$ cannot be. \\

{\bf 5 -} We may now write explicit formulae for $\mu_{2k}$ with $0
\leq k \leq 2n-1$~:

$$\mu_{2k} = \sum_{i=1}^n \lambda_i \ x_i^k \qquad (0 \leq k \leq 2n-1)
\eqno({\rm C.18})$$

\noi where $x_1, \cdots , x_n$ are the roots of $p_n$, and the
coefficients $\lambda_i$ are given by~:

$$\lambda_i = {1 \over p'_n(x_i)} \ <{p_n(x) \over x - x_i}>\ . 
\eqno({\rm C.19})$$

\noi To prove (C.18), notice that it amounts to~:

$$<P> \ = \sum_{i=1}^n \lambda_i \ P(x_i) \qquad \hbox{(any $P$ of 
degree $\leq 2n-1$)}\ . \eqno({\rm C.20})$$

\noi Performing the Euclidean division of $P$ by $p_n$, we have

$$P = Qp_n + R \eqno({\rm C.21})$$

\noi with degree $Q < n$ and degree $R < n$. We may verify (C.20)
separately for $Qp_n$ and for $R$. \par

For $P = Qp_n$, the left-hand side of (C.20) vanishes by (C.16), and
the right-hand side vanishes because the $x_i$ are the roots of $p_n$.
\par

For $P = R$, we use the identity

$$R(x) = \sum_{i=0}^n {1 \over p'_n(x_i)} \ {p_n(x) \over x - x_i} \
R(x_i) \eqno({\rm C.22})$$

\noi which stems from the fact that both sides are polynomials of degree
$< n$, that are equal at $n$ points $x=x_i$. Then (C.20) is satisfied due to
the choice (C.19) of the coefficients $\lambda_i$. \\

\noi {\bf 6 -} Define

$$\rho_n (x) = \sum_{i=1}^n \lambda_i \ \delta (x - x_i) \ .\eqno({\rm C.23})$$

\noi Then (C.18) writes

$$\mu_{2k} = \int_0^{\infty} \rho_n (x) \ x^k \ dx \qquad (0 \leq k
\leq 2n-1) \eqno({\rm C.24})$$

\noi to be compared with formula (C.6) to be proved, which writes

$$\mu_{2k} = \int_0^{\infty} \rho_N (x) \ x^k \ dx \qquad (0 \leq k
\leq N) \eqno({\rm C.25})$$

\noi with

$$\rho_N(x) = 2 \pi \ \sqrt{x} \ \left | \psi_0^{(N)}(\sqrt{x})\right
|^2 \ . \eqno({\rm C.26})$$

\noi We have still the problem that $\rho_n(x)$ is not a function.\par

The idea to solve this problem is to vary $\mu_{4n-2}$, keeping fixed
$\mu_{2k}$ for $0 \leq k \leq 2n-2$. The polynomial $p_n$ then depends
on $\mu_{4n-2}$ as a parameter, and as well its zeros $x_i(\mu_{4n-2})$
and the coefficients $\lambda_i(\mu_{4n-2})$ defined by (C.19). Then
formula (C.24) is lost for $\mu_{4n-2}$, but remains valid for $0 \leq
k \leq 2n-2$, and in fact gives a whole family of formulae

$$\mu_{2k} = \int_0^{\infty} \rho_n (\mu_{4n-2}, x) \ x^k\ dx \qquad (0 \leq k
\leq 2n-2)\eqno({\rm C.27})$$

\noi with a weight distribution

$$\rho_n(\mu_{4n-2}, x) = \sum_{i=1}^n \lambda_i \left ( \mu_{4n-2})\
\delta (x - x_i(\mu_{4n-2}) \right ) \eqno({\rm C.28})$$

\noi depending on $\mu_{4n-2}$. We may then take the mean value of
(C.27) over any interval $[\mu_{4n-2}^{(1)}, \mu_{4n-2}^{(2)}]$ in
which the constraints are satisfied, obtaining

$$\mu_{2k} = \int_0^{\infty} \rho_n(x) \ x^k \ dx \qquad (0 \leq k \leq
2n-2) \eqno({\rm C.29})$$

\noi with

$$\rho_n(x) = {1 \over \mu_{4n-2}^{(2)} - \mu_{4n-2}^{(1)}}
\int_{\mu_{4n-2}^{(1)}}^{\mu_{4n-2}^{(2)}} \rho_n (\mu_{4n-2}, x)
d\mu_{4n-2} \ . \eqno({\rm C.30})$$

Now, $\rho_n(x)$ defined by (C.30) has a good chance to be a genuine
function, because integrating a $\delta$ distribution over a parameter
usually gives a function. \par

However, there is an obvious case in which this does not hold, namely
when the point where the $\delta$ distribution is concentrated does not
depend on the parameter. So we still have to show that {\it each} zero
of $p_n$ does vary with $\mu_{4n-2}$. Let us consider the orthogonal
polynomials $\widetilde{p}_0, \widetilde{p}_1, \cdots$ with respect to
the new scalar product $\widetilde{<P|P'>}$ associated to new values
$\widetilde{\mu}_0, \widetilde{\mu}_2 , \cdots$ of the moments, with
$\widetilde{\mu}_{2k} = \mu_{2k}$ for $0 \leq k \leq 2n-2$, and
$\widetilde{\mu}_{4n-2} \not= \mu_{4n-2}$. Note that the new scalar
product of two polynomials is the same as the original one when the sum
of the degrees is $\leq 2n-2$. It follows that $\widetilde{p}_k = p_k$
for $0 \leq k \leq n-1$, and also that

$$<\widetilde{p}_n|p_k> = < \widetilde{p}_n|\widetilde{p}_k> = 0 \qquad
\hbox{for \quad $0 \leq k \leq n-2$}\ . \eqno({\rm C.31})$$

\noi Therefore, the expansion of $\widetilde{p}_n$ over the $p_k$
writes~:

$$\widetilde{p}_n = p_n + c\ p_{n-1} \ . \eqno({\rm C.32})$$

\noi And one has $c \not= 0$. Indeed, since $\widetilde{
<\widetilde{p}_n|\widetilde{p}_{n-1}>} = 0$, one has

$$c = {1 \over <p_{n-1}|p_{n-1}>} \left ( < \widetilde{p}_n | p_{n-1}>
- \widetilde{<\widetilde{p}_n | \widetilde{p}_{n-1}>} \right ) $$
$$ = {1 \over <p_{n-1}|p_{n-1}>} \left ( <\chi^n|\chi^{n-1} >\ - \ 
\widetilde{<\chi^n|\chi^{n-1}>} \right )\eqno({\rm C.33})$$

\noi so that

$$c =  {1 \over <p_{n-1}|p_{n-1}>} \left (  \mu_{4n+2} -
\widetilde{\mu}_{4n-2}  \right )\ . \eqno({\rm C.34})$$
\vskip 3 truemm

The fact that a zero of $p_n$ cannot be a zero of $\widetilde{p}_n$ now
follows from (D.32) and the fact that a zero of $p_n$ cannot be a zero
of $p_{n-1}$. \par

This last point is a well known property of orthogonal polynomials,
which can be proved directly as follows. Assume that a zero $x_i$ of
$p_n$ is also a zero of $p_{n-1}$. Then we have

$$<p_{n-1} | {p_n \over x - x_i}> \ = \ <{p_{n-1} \over x - 
x_i}|p_n > \ = 0 \eqno({\rm C.35})$$

\noi by (C.16). On the other hand, writing ${p_n \over x - x_i} = a
x^{n-1} + \cdots$, where $a \not= 0$, we have again by (C.16)~:

$$<p_{n-1}|{p_n \over x - x_i}> \ = \ a <p_{n-1}|x^{n-1}> \ = \ a
<p_{n-1}|p_{n-1}> \eqno({\rm C.36})$$

\noi which cannot vanish, contradicting (C.35).\par

We are now in position to complete the proof of (C.6). Indeed, using
the implicit functions theorem, one can infer from (C.32) and (C.34)
that, for a small enough interval $[\mu_{4n-2}^{(1)},
\mu_{4n-2}^{(2)}]$, each function $x_i(\mu_{4n-2})$ is a diffeomorphism
of this interval to an interval $[x_i^{(1)}, x_i^{(2)}]$ in $x$. Then
introducing the reciprocal function $x_i \to \mu_i(x_i)$ of $\mu_{4n-2}
\to x_i(\mu_{4n-2})$, the integral of a $\delta$ function is computed
by changing the variable of integration $\mu_{4n-2}$ to $x_i =
x_i(\mu_{4n-2})$~:

$$\int_{\mu_{4n-2}^{(1)}}^{\mu_{4n-2}^{(2)}} \lambda_i \left (
\mu_{4n-2}\right ) \ \delta \left ( x - x_i\left ( \mu_{4n-2}\right )
\right ) \ d\mu_{4n-2} = \int_{x_i^{(1)}}^{x_i^{(2)}} \lambda_i \left (
\mu_i(x_i)\right ) \ \delta (x-x_i) \ |\mu'_i(x_i)| dx_i$$ $$ =
\chi_{[x_i^{(1)}, x_i^{(2)}]}(x) \ \lambda_i\left ( \mu_i(x)\right ) \
|\mu'_i(x)| \eqno({\rm C.37})$$

\noi where $\chi_I$ is the characteristic function of an interval $I$,
namely $\chi_I(x) = 1$ for $x \in I$ and $\chi_I(x) = 0$ for $x \notin
I$. Then (C.30) gives

$$\rho_n(x) = {1 \over \mu_{4n-2}^{(2)} - \mu_{4n-2}^{(1)}} \ 
\sum_{i=1}^n \ \chi_{[x_i^{(1)},
x_i^{(2)}]}(x)\ \lambda_i \left ( \mu_i(x\right )\ |\mu'_i(x)|
\eqno({\rm C.38})$$

\noi and this is a genuine positive function, which can therefore be
written as the square of a wave function.\par

\vskip 1 truecm

\section*{Appendix D. Explicit lower limits for the even derivatives.}
\hspace*{\parindent} In this Appendix we generalize to all even
derivatives $\xi^{(2n)}(1)$ the proof of the bound (\ref{73e}) that we
have given in ref. \cite{3r}.

 From expression (\ref{43e})~:

$$
\xi_{NR}^{(m)}(1) = (-1)^m\ 2^m\ {m! \over (2m)!} \ \left (
m_q\right )^{2m} \ <0|z^{2m}|0> \eqno({\rm D.1})$$

\noi using rotational invariance we obtain

$$
\xi_{NR}^{(m)}(1) =  (-1)^m\ {m! \over (2m)!} \ \left (
m_q\right )^{2m} \ {1 \over 2m+1} \ <0|r^{2m}|0> \eqno({\rm D.2})$$

\noi and from

$$<0|r^{2m}|0> \ = \left | <0|r^{m}|0>\right |^2 +
\sum_{n\not= 0} \left | <n|r^m|0>\right |^2 \ . \eqno({\rm D.3})$$

\noi Using again spherical symmetry

$$<0|r^{2m}|0> \ = (m+1)^2 \left | <0|z^{m}|0>\right |^2
+ (m+1)^2 \sum_{n\not= 0,rad} \left | <n|z^m|0>\right |^2 \eqno({\rm D.4})$$

\noi one obtains
$$\xi_{NR}^{(m)}(1) =  (-1)^m\ {m! \over (2m)!} \
\left ( m_q\right )^{2m} \ {(m+1)^2 \over 2m+1}$$
$$\left [ \left
| <0|z^{2m}|0>\right |^2 + \sum_{n\not= 0,rad} \left | <n|z^m|0>\right
|^2\right ]\eqno({\rm D.5})$$

\noi and therefore

$$(-1)^m\ \xi_{NR}^{(m)}(1) > {m! \over (2m)!} \ \left (
m_q\right )^{2m} \ {(m+1)^2 \over 2m+1}  \ \left | <0|z^m|0>\right |^2
\ . \eqno({\rm D.6})$$

This expression demonstrates that $\xi_{NR}(w)$ is an alternate series
in powers of $(w-1)$. \par

Assuming $m$ to be even, $m = 2n$, one gets

$$\xi_{NR}^{(2n)}(1) > {(2n)! \over (4n)!} \ \left (
m_q\right )^{4n} \ {(2n+1)^2 \over 4n+1}  \ \left | <0|z^{2n}|0>\right
|^2 \ . \eqno({\rm D.7})$$

\noi The moment $<0|z^{2n}|0>$ can be expressed in terms of
$\xi_{NR}^{(n)}(1)$ though (D.1), giving finally

$$\xi_{NR}^{(2n)}(1) >  {[(2n)!]^3 \over [n!]^2
(4n)!} \ {(2n+1)^2 \over 4n+1} \ \left [\xi_{NR}^{(n)}(1) \right ]^2
\qquad (n \geq 0) \ . \eqno({\rm D.8})$$

\noi We obtain, for the lower values of $n$,
$$n = 1 \qquad \xi_{NR}^{(2)}(1) \geq {3 \over 5}  \left
[\xi_{NR}^{(1)}(1) \right ]^2 \ \ \ $$
$$n=2 \qquad
\xi_{NR}^{(4)}(1) \geq {5 \over 21}  \left [\xi_{NR}^{(2)}(1) \right
]^2 \ . \eqno({\rm D.9})$$

\noi The formula (D.8) generalizes the result (\ref{73e}) to all
even derivatives. \par

For the odd derivatives one gets, with the present method, from 
(D.6), the weaker result

$$- \xi_{NR}^{(2n+1)}(1) > 0 \qquad (n \geq 0) \ . \eqno({\rm D.10})$$

We see that we had obtained in Sections 3 and 4 a much stronger result
for $\xi_{NR}^{(4)}(1)$ than (D.9), and non trivial results for $-
\xi_{NR}^{(3)}(1)$ and $- \xi_{NR}^{(5)}(1)$. However, we have obtained
here an explicit lower bound for $\xi_{NR}^{(2n)}(1)$ (D.8).

\section*{Acknowledgement.}

We acknowledge support from the EC contract HPRN-CT-2002-00311.

\end{document}